\documentstyle[aps,prb,psfig]{revtex}
\begin{document}

\title{Thermal conductivity in the vortex state of YBa$_2$Cu$_3$O$_7$
and Sr$_2$RuO$_4$}
\author{L. Tewordt and D. Fay}
\address{I. Institut f\"ur Theoretische Physik,
Universit\"at Hamburg, Jungiusstr. 9, 20355 Hamburg, 
Germany}
\date{\today}
\maketitle
\begin{abstract}
      The thermal conductivity $\kappa\,$ is calculated for the Abrikosov vortex
lattice with d-wave pairing. The Andreev scattering rate arises from the self
energy term in Gorkov's integral equation for the corresponding Green's
function. The impurity scattering rate is calculated self-consistently from
this Green's function in the t-matrix approximation. Neither point-like Impurity
scattering with a phase shift in the almost unitary limit nor in the Born limit can
consistently explain recent zero field and finite field measurements of 
$\kappa/T$ at low $T$ in ultraclean YBCO.  Our theory also applies to the spin 
triplet pairing states with vertical or horizontal line nodes in Sr$_2$RuO$_4\,$.
\end{abstract}
\vspace{0.5in}
        Recently the thermal conductivity $\kappa\,$ of ultraclean 
YBa$_2$Cu$_3$O$_7$ was measured at very low temperatures in magnetic
fields up to 13 Tesla. \cite{Hill} In zero field, $\kappa/T\,$ rises from its
universal limit $\kappa_{00}/T\,$ very fast as a function of temperature which,
according to the quasiclassical calculation for a d-wave superconductor,\, 
\cite{Graf} indicates either an extremely small normal state impurity scattering
rate $\Gamma=1/2\tau$ in the unitary limit (phase shift $\rho = \pi/2$), or an
unrealistically large $\Gamma$ for the Born approximation 
(phase shift $\rho = 0\,$). As a function of field, $\kappa\,$ initially increases
very rapidly reaching nearly twice its universal limit $\kappa_{00}\,$ and
remains almost unchanged up to 13 Tesla. The theory of Vekhter and
Houghton, \cite{Vek}  which includes scattering of quasiparticles at the 
vortices via Andreev scattering, yields a rise proportional to $\sqrt{H}$
in agreement with the measurements for a less pure sample of YBCO. 
\cite{Chiao} Recently the data of Ref.~\onlinecite{Hill} have been accounted
for by assuming a phase shift $\rho$ for impurity scattering slightly less than
the unitary limit, $\pi/2\,$, and adding to the extremely small impurity
scattering rate $\Gamma$ a phenomenological quasiparticle scattering due
to Andreev scattering at the superfliud flow of the vorticies. \cite{Kim}

      In the present paper we use the theory of Ref.~\onlinecite{Vek} and 
the equivalent theory of Ref.~\onlinecite{TF} to solve the problem
of the occurence of a plateau-like feature in the field dependence of 
$\kappa\,$ in YBCO. These theories are based on the theory of Brandt, 
Pesch, and Tewordt (BPT-theory)\, \cite{BPT} where the spatial average of
the Gorkov Green's function $G$ for the Abrikosov vortex lattice was 
calculated. The expression for the thermal conductivity which has been 
derived from this Green's function \cite{TF} can explain the measured field
dependence of $\kappa\,$ for different field directions for superconducting
spin-triplet pairing states with vertical or horizontal nodes in 
Sr$_2$RuO$_4\,$. A simplified version of the BPT-theory which is based on
the quasiclassical Green's function $g\,$, \,\cite{Pesch,Klim} referred to here 
as the P-approximation, yields the expression of Ref.~\onlinecite{Vek} for the
thermal conductivity $\kappa \equiv \kappa_{xx}$ in the vortex state with
the field perpendicular to the basel plane for a d-wave superconductor. We
find that the expressions for the thermal conductivity derived from the
BPT-Gorkov or from the P-quasiclassical Green's functions 
(Refs.~\onlinecite{TF} and ~\onlinecite{Vek}) yield to a very good
approximation the same results for all reduced fields $h=H/H_{c2}\,$, 
impurity scattering rates $\delta=\Gamma/\Delta_0\,$, and phase 
shifts $\rho\,$. Here $\Delta_0$ is the amplitude of the d-wave order 
parameter.

    To save space we omit here the full expression for $\kappa\,$ and
concentrate on the discussion of the Andreev scattering which is derived
from first principles. This contribution to the quasiparticle scattering is
contained in the expression for Im$\xi_0$ in the denominator of the 
$\omega$-integral for $\kappa$ which also includes the well-known factor
$\omega^2\, \mbox{sech}^{2}(\omega/2T)\,$.  \cite{TF} Here $\xi_0$ is the
position of the pole of the BPT-Green's function $G$ as a function of the
normal state energy $\xi$ measured from the Fermi energy. The equation
for the zero $\xi_0$ of the denominator of $G$ yields 
Im$\xi_0 = \gamma_i + \gamma_A\,$, where $\gamma_i$ is the total
impurity scattering rate and $\gamma_A$ is the imaginary part of the
quasiparticle energy $-\Sigma_A$ at $\xi_0\,$. The kernel of Gorkov's 
integral equation for $G$ in the spatial representation yields
\begin{equation}
\Sigma_A(r_1,r_2;\omega)=-\Delta(r_1)\Delta^\ast(r_2)G^0(r_1-r_2;-\omega)\, ,
\label{Sigr}
\end{equation}
where $\Delta(r)$ is Abrikosov's vortex lattice order parameter. From this
expression it is clear that $\gamma_A = -\mbox{Im}\Sigma_A$ is the
scattering rate for converting a quasiparticle at $\Delta^{\ast}(r_2)$ by
Andreev reflection into a quasihole and then back into a quasiparticle at
$\Delta(r_1)\,$. The Fourier transform of $\Sigma_A(r_1-r_2;\omega)$
yields\, \cite{BPT}
\begin{equation}
\Sigma_A(\mbox{\boldmath$p$} ,\omega)=-i\sqrt{\pi}\Delta^2(\Lambda/v_{\perp})
w[(\omega+i\gamma_i+\xi_p)\Lambda/v_{\perp}]\, .
\label{Sigp}
\end{equation}
Here $\Delta^2$ is the spatial average of $|\Delta(r)|^2\,$, 
$\Lambda=(2eH)^{-1/2}$ is of the order of the vortex lattice constant, 
$v_{\perp}(\mbox{\boldmath$p$})$ is the Fermi velocity component 
perpendicular to the
applied field $\mbox{\boldmath$H$}\,$, and $w$ is Dawson's integral. In the limit
$\omega \rightarrow 0$ corresponding to the $T \rightarrow 0$ limit of 
$\kappa/T\,$, one obtains explicit expressions for 
Im$\xi_0 = \gamma_i + \gamma_A(\xi_0)$ and $\kappa/\kappa_n\,$.
\cite{TF} In the limit $v_{\perp} \rightarrow 0\,$, i.e., 
$\mbox{\boldmath$v$}(\mbox{\boldmath$p$}) \parallel \mbox{\boldmath$H$}\,$, 
Eq.(\ref{Sigp}) tends to the
self energy of the BCS Green's function. In the limit 
$\Lambda \rightarrow \infty\,$, i.e., $H \rightarrow 0\,$, the expressions for 
$\kappa$ in the BPT-approximation \cite{TF} and in the equivalent 
P-approximation \cite{Vek} tend correctly to the expression for $\kappa$
which was first derived in Ref.~\onlinecite{AmbTew}.

     In the following we present results for a d-wave pairing state in a field
perpendicular to the basel plane where the Abrikosov vortex lattice state is
multiplied by $\cos(2\phi)\,$. The impurity scattering rate $\gamma_i$ is
calculated self-consistently in the t-matrix approximation for the self
energy $\Sigma_i\,$:
\begin{equation}
\gamma_i = \mbox{Re}\Sigma_i\: ; 
\qquad \Sigma_i = \Gamma\,\bar{g}(\omega)
\left[ \cos^2\rho + \bar{g}^2\,\sin^2\rho \right]^{-1}\, ;
\label{Gam}
\end{equation}
\begin{equation}
g(\omega,\phi) = \left[ 1 - i\sqrt{\pi}(2\Delta\Lambda/v)^2
\cos^2(2\phi)w^{\prime}(z) \right]^{-1/2}\, ;
\label{g}
\end{equation}
\begin{equation}
z = 2 (\omega + i\,\Sigma_i)\Lambda/v\: ;
\qquad \Lambda=(2eH)^{-1/2}\: .
\label{z}
\end{equation}
Here $g$ is the quasi-classical Green's function in the P-approximation,
\cite{Pesch} and $\bar{g}$ is the angular average of $g\,$. The field
dependence of $\Delta$ is approximately given by 
$\Delta = \Delta_0 \sqrt{1-h}\,$, where $h=H/H_{c2}\,$.

      In Fig.1(a) we show the ratio of $\kappa_{xx} \equiv \kappa$ to the normal
state conductivity $\kappa_n$ vs $h=H/H_{c2}$ for $\omega=0$ 
corresponding to the limit $T \rightarrow 0\,$, and the density of states 
$N/N_0 = \mbox{Re}\,\bar{g}(0)\,$, for reduced normal state scattering rates
$\delta = \Gamma/\Delta_0 = \,$ 0.1, 0.01, 0.001, and 0.0001 in the Born
approximation (phase shift $\rho=0$). The occurence of a plateau in the low
field range for all scattering rates is demonstrated more clearly in Fig.1(b).
In Fig.2(a) we show our corresponding results for $\kappa/\kappa_n$ and
$N/N_0$ for a phase shift $\rho = 0.495\pi$ which is very close to the unitary
limit $\rho=\pi/2$ and has been used in Ref.~\onlinecite{Kim} to fit the data of
Ref.~\onlinecite{Hill}. Fig.2(b) shows more clearly that a plateau in the low 
field range is reached only approximately in the limit of very small scattering
rates, here $\delta=0.001$ and 0,0001. The latter value is of the order of
magnitude of the value that has been used in Ref.~\onlinecite{Kim} to fit the
data of Ref.~\onlinecite{Hill}. We find that the field dependence of 
$\kappa/\kappa_n$ is very nearly the same as the field dependence of the
ratio of the scattering rates in the normal and vortex states, i.e., the angular
average of $\Gamma/(\gamma_i + \gamma_A)\,$. This means, according to
Figs.1(a) and 2(a), that the Andreev scattering rate $\gamma_A$ is much
larger than the impurity scattering $\Gamma$ up to fields just below
$H_{c2}\,$, and that the ratio $\gamma_A/\Gamma$ at fixed $h$ increases
with decreasing scattering rate $\delta\,$.

      We now attempt to explain the second feature of the experiments of 
Ref.~\onlinecite{Hill}, i.e., that $\kappa/T$ becomes almost temperature
independent up to about 0.6 K for constant fields above 1 Tesla up to 13 
Tesla. In Fig.1(c) we show our results for $\kappa(\omega)/\kappa_n$ vs 
$\,\Omega=\omega/\Delta_0\,$ for $\delta=0.1$ and 0.05 and several fields 
$h=$ 0.1, 0.06, and 0.02 in the Born approximation ($\rho=0$). Here, 
$\kappa(\omega)/\kappa_n$ is the factor multiplying 
$(\omega/T)^2 \,\mbox{sech}^{2}(\omega/2T)$ in the normalized integral 
over $d(\omega/T)$ for $\kappa/\kappa_n\,$. Since 
$(\omega/T)^2 \,\mbox{sech}^{2}(\omega/2T)$ has a maximum at 
$\omega/T = 2.4\,$, one has
$\Omega = \omega/\Delta_0 \simeq 2.4(T/\Delta_0)\,$, and thus  
$\kappa(\omega)/\kappa_n$ yields approximately the dependence of
$\kappa/\kappa_n \propto \kappa/T$ as a function of 
$\Omega \simeq 2.4(T/\Delta_0)\,$. We have divided 
$\kappa(\omega)/\kappa_n$ by $\delta$ because 
$\kappa_n\delta \sim (\pi/2)\kappa_{00}$ where $\kappa_{00}$ is the
universal conductivity limit. One recognizes from Fig.1(c) that 
$\kappa(\omega)/\kappa_n\delta$ is almost constant in the range from
$\Omega = 0$ to 0.1 which means that $\kappa/T$ is almost constant
up to about $T/T_c \sim 0.1\,$, and that the curves for constant fields
$h=$ 0.1, 0.06, and 0.02 lie close together for each of the two scattering
rates $\delta=0.1$ and 0.05. In Fig.2(c) we show the corresponding results
for the phase shift $\rho = 0.495\pi$ and scattering rate $\delta = 0.0001\,$,
and in Fig.3(a) the results for scattering rate $\delta = 0.001$ and the same
phase shift and fields. Again these results indicate that $\kappa/T$ is nearly
constant as a function of $T$ at least for the higher fields $h=0.1$ and 0.06.
However, these curves for the almost unitary limit are not as close to each
other as those for the Born approximation in Fig.1(c). This is in disagreement
with the data for the temperature dependence of $\kappa/T$ for fixed fields
between 0.8 and 13 Tesla which lie on about the same level. \cite{Hill}

     We have also calculated $\kappa(\omega)/\kappa_n\delta$ in the limit of
zero field. The expression for $\kappa$ in Ref.~\onlinecite{TF} yields in the
limit $H\rightarrow 0\,$, or $\Lambda \rightarrow \infty\,$, the following
result:
\begin{equation}
\frac{\kappa(\omega)}{\kappa_n\delta} 
=\int^{2\pi}_0 \frac{d\phi}{\pi} \,\cos^2(\phi) \,
\frac{1}{\mbox{Im}\left[ \tilde{\Omega}^2 - \cos^2(2\phi) \right]^{1/2}}
\,\frac{1}{2} \left( 1 + \frac{|\tilde{\Omega}|^2 - \cos^2(2\phi)}
{|\tilde{\Omega}^2 - \cos^2(2\phi)|} \right)\, ;
\label{kappa}
\end{equation}
\begin{equation}
\tilde{\Omega} = \Omega + i\,\Sigma_i/\Delta_0\: ;
\qquad g(\tilde{\Omega},\phi) = \tilde{\Omega} 
\left[ \tilde{\Omega}^2 - \cos^2(2\phi) \right]^{-1/2}\: ;
\qquad \Omega=\omega/\Delta_0\: .
\label{Omega}
\end{equation}
The impurity self energy $\Sigma_i$ is calculated self-consistently with
the help of Eq.(\ref{Gam}). Eq.(\ref{kappa}) agrees with the general 
strong-coupling result of Ref.~\onlinecite{AmbTew}. In Fig.2(c) we have
plotted our results for $\kappa(\omega)/\kappa_n\delta$ for phase shift
$\rho=0.495\pi$ and scattering rate $\delta=0.0001\,$, and in Fig.3(a) we
show the results for the same phase shift and scattering rates 
$\delta=$0.001, 0.01, and 0.1. One sees from these figures that for 
$H=0$ the ratio 
$\kappa(\omega)/\kappa_n\delta \simeq (2/\pi)\kappa(\omega)/\kappa_{00}$
tends for $\Omega\rightarrow0\,$, or $T\rightarrow0\,$, correctly to the
value $(2/\pi)\,$, and that this function rises approximately proportional to
$\Omega^2$ with a slope that decreases for increasing $\delta\,$. The zero 
field and constant field curves in Figs.2(c) and 3(a) for $\delta=$0.0001 and 
0.001 are qualitatively similar to the data for the $T$ dependence of $\kappa/T$
in zero and constant fields. \cite{Hill} However, the values of $\kappa/T$ in
units of $\kappa_{00}/T$ shown in Fig.2(c) for the constant fields $h=$0.1, 
0.06, and 0.02 are an order of magnitude too large in comparison to the 
experimental values. The corresponding values for $\delta=0.001$ shown in
Fig.3(a) are much smaller. These values can be reduced further to reach the
experimental value of about 2 with a reasonable increase in the factor 
$\alpha$ multiplying the quantity $\Delta\Lambda/v$ [see Eqs. (\ref{Sigp}) and
(\ref{g})] for a d-wave pairing state. \cite{Vek} Here $\alpha\propto v_2/v$ 
where $v_2$ is the slope of the d-wave gap at the node. The dashed curves
in Fig.3(a) correspond to the value $\alpha=5\,$. These curves are
reduced to those in Fig.3(b) by using the value $\alpha=31\,$. The upper
dashed curves in Fig.2(c) for $\delta=0.0001$ also correspond to $\alpha=5$
and they are reduced to the lower dashes curves by taking the same value 
of $\alpha=31$ as for $\delta=0.001$ in Fig.3(b). A still much larger value of 
$\alpha$ is needed to reach the experimental value
$\kappa/\kappa_{00} \sim 2\,$. That corresponds to an unrealistically large
value of $v_2/v\,$, We thus conclude that the measured $T$ dependence of 
$\kappa/T$ for ultraclean YBCO in
zero field, and finite fields up to 13 Tesla, can be qualitatively explained 
by our theory with the parameter values $\rho=0.495\pi$ and $\delta=0.001$
and 0.0001.
However, the corresponding curves for the field dependence of 
$\kappa/\kappa_n$ in the $T \rightarrow 0$ limit [see Fig.2(b)] do not rise
as abruptly to the plateau-like value for increasing field as the experimental
curve. Contrary to the nearly unitary limit, in the Born 
approximation the field dependence of $\kappa/\kappa_n$ shown in 
Fig.1(b), and the temperature dependence of $\kappa/T\,$ [see Fig.1(c)], are,
for example for $\delta=0.1\,$, in much better agreement with the 
experiments of  Ref.~\onlinecite{Hill}. However, the $T$ dependence of
$\kappa/T$ for zero field is in total disagreement with the data: it rises
steeply for very small $\Omega$ and then tends to a constant value of
about $4\kappa_{00}/T$ for $\Omega$ up to about 0.1 [see Fig.1(c)].

     We have also investigated intermediate values of the impurity 
scattering phase shift, for example, $\rho=0.4\pi$ which has been used to
fit the data for the microwave conductivity in YBCO films. \cite{Hensen} We
find that the measured field dependence of the thermal conductivity \cite{Hill}
is better described with $\rho=0.4\pi$ and $\delta=0.001$ or 0.0001 than with
$\rho=0.495\pi$ and and the same $\delta$'s. However, the $\Omega$
dependence of $\kappa/\kappa_{00}$ for zero field is quite far from a 
$\Omega^2$ dependence: it rises steeply to a maximum and then decreases
slowly for increasing $\Omega\,$. Thus phase shifts intermediate
between the Born approximation and the unitary limit do not seem useful for
explaining the experiments of Ref.~\onlinecite{Hill}.

     The aim of the theory for $\kappa$ in Ref.~\onlinecite{TF} was to explain
the measurements of $\kappa$ in Sr$_2$RuO$_4$ for fields perpendicular
to the ab plane and for rotating in-plane fields. \cite{Izawa} Recently, 
$\kappa$ was measured in Sr$_2$RuO$_4$ in zero field down to very low
temperatures. \cite{Suzuki} This low-temperature behaivior of $\kappa$
demonstrates the universal character of the heat transport due to a gap 
with nodes and suggests strong impurity scattering with a phase shift close
to $\pi/2\,$. Since our present theory of $\kappa$ for a d-wave
superconductor also applies to a spin triplet f-wave pairing state with
vertical or horizontal line nodes, \cite{TF} we conclude that our results for
phase shifts $\rho=0.495\pi \simeq \pi/2$ also apply to Sr$_2$RuO$_4ß,$.
The $T$ dependence of $\kappa/T$ for $\delta=0.01$ and 0.1 shown in
Fig.3(a) for zero field is similar to the data in Ref.~\onlinecite{Suzuki}. The
curves for $\kappa/\kappa_n$ vs $H/H_{c2}$ for $\delta=0.01$ ansd 0.1 
in Fig.2(a) lie far below the curves for $\delta=0.5$ and 0.2 obtained in
the Born approximation in Ref.~\onlinecite{TF}.

     In summary, we have calculated the thermal conductivity for the 
Abrikosov vortex lattice state with d-wave pairing which automatically
includes Andreev scattering due to the self energy in Gorkov's integral
equation for the Green's function. The most important result of the present
paper is that the BPT-approximation \cite{BPT,TF} based on the Gorkov
equations yields very nearly the same results for $\kappa$ as the 
P-approximation \cite{Klim,Vek} based on the equations for the quasiclassical
Green's functions. These results for $\kappa$ correspond to previous results 
for the density of states $N/N_0$ for d-wave pairing in the vortex state. 
\cite{Dahm} In Ref.~\onlinecite{Dahm} it was shown that the BPT- and 
P-approximations yield nearly the same results as the solutions of the
quasiclassical equations for applied fields between $H_{c2}$ and 
$H_{c1}$ while the Doppler shift approximation gives rise to substantial 
errors. In agreement with Ref.~\onlinecite{Vek} we find that, for impurity
scattering in the Born limit, $\kappa$ exhibits a steep initial increase as a
function of field and then becomes almost constant (see Fig.1) while, in the
almost unitary limit, $\kappa$ rises like $\sqrt{H}$ (see Fig.2). The latter 
result disagrees with the results of Ref.~\onlinecite{Kim} where, in the almost
unitary limit, $\kappa$ rises initially very rapidly and then becomes almost
constant with a value in units of the universal conductivity $\kappa_{00}$ close
to the experimantal value. The reasons for the discrepency are apparently the
following. First, the renormalization of the impurity scattering self energy in the 
t-matrix approximation is carried out in Ref.~\onlinecite{Kim} by employing the
zero-field density of states while, in Ref.~\onlinecite{Vek} and here 
[see Eqs.(\ref{Gam}) and (\ref{g})], the density of states is calculated 
self-consistently for finite field. Second, the Andreev scattering rate 
$\gamma_A$ is approximated in Ref.~\onlinecite{Kim} by $b\,E_H$ where 
$E_H\sim v/\Lambda\,$ [see Eq.(\ref{z})] is the magnetic field energy and $b$ is
a parameter which is fitted to the data of Ref.~\onlinecite{Hill}. This expression
for the Andreev scattering rate is taken in analogy to the expression derived
by the method of the Doppler shift followed by an averaging over a single 
vortex. \cite{Yu}  However, the expression for $\gamma_A$ derived in 
Ref.~\onlinecite{Yu} contains an additional exponential factor which changes the 
field dependence considerably. In our approach 
$\gamma_A = -\mbox{Im}\Sigma_A$ where $\Sigma_A$ is the self energy in the
Gorkov equation which has the form for Andreev scattering by the spatial 
variation of the complex order parameter $\Delta(r)$ of the total Abrikosov 
vortex lattice [see Eq.(\ref{Sigr})]. It is interesting that the exponential function
arising from the imaginary part of Eq.(\ref{Sigp}) has, in the absence of impurity
scattering $\gamma_i\,$, a similar form as that of Ref.~\onlinecite{Yu} 
considered as a function of frequency, field, and angle 
$\theta=\angle(\mathbf{p},$\boldmath$H$\unboldmath$)\,$. Our $\gamma_A$ in 
the absence of impurity scattering is shown as a function of $\theta$ for 
several different fields and frequencies in Fig.5 of Ref.~\onlinecite{TF2}. It should
be noted that here $\gamma_A$ depends also on $\gamma_i$ 
[see Eq.(\ref{Sigp})] where both quantities are calculated self-consistently 
together with Eqs.(\ref{Gam}) - (\ref{z}). Although the theory of 
Ref.~\onlinecite{BPT} was originally derived for applied fields $H$ near 
$H_{c2}\,$, it has been shown to work well over the entire range of linear 
magnetization, \cite{Brandt} and it has been tested down to $H_{c1}$ for a
d-wave superconductor by comparison with the solutions of the quasiclassical
equations. \cite{Dahm}

     Let us now briefly discuss why our theory in its present form fails to explain
the experiments of Ref.~\onlinecite{Hill} consistently for zero and finite fields.
The measured rapid growth of the zero field $\kappa/T$ with $T$ in ultraclean 
YBCO can, in agreement with Ref.~\onlinecite{Kim}, indeed be explained only 
by assuming a very small impurity scattering rate 
$\delta=\Gamma/\Delta_0 \sim 10^{-3}$ to $10^{-4}$ and a phase shift close
to the unitary limit $\pi/2$ for isotropic scatterers. The observed sudden onset 
of a "plateau" in $\kappa/T$ as a function of field $H$ requires, for this phase
shift limit, the assumption of a very small $\delta\le10^{-3}$ because this makes
the region of $\sqrt{H}$-behavior small and $\kappa$ beyond this region
nearly constant as a function of $H\,$. The third feature of the experiments, 
i.e., that $\kappa/T$ is independent of $T\,$, is also satisfied. However, the
saturation values of $\kappa/T$ in units of $\kappa_{00}/T$ are too large and 
lie too far apart for different fields in comparison to the experimental values. 
The Born approximation (phase shift zero) with $\delta \sim 0.1$ to $0.01$
yields a much better description of the measured field dependence of 
$\kappa\,$, i.e., a steep initial increase followed by a plateau, and the
saturation values of $\kappa/T$ in units of $\kappa_{00}/T$ for different fields
are in fair agreement with the data. However, the calculated $T$ dependence
of $\kappa/T$ in zero field is quite different from the observed temperature 
dependence if one assumes isotropic impurity scattering in the Born limit. 
This is also the case for intermediate phase shifts between $\rho=0$ and
$\rho=\pi/2\,$, for example for $\rho=0.4\pi$ which has been used to fit 
microwave conductivity data. \cite{Hensen}

     In conclusion, we have shown that the measured field dependence of 
$\kappa/T$ in ultraclean YBCO, \cite{Hill} in particular the plateau, can be 
better described by the Born approximation than by the almost unitary
phase shift limit for impurity scattering. The use of the almost unitary limit
is however necessary \, \cite{Kim} in order to explain the observed rapid 
increase of $\kappa/T$ with $T$ in zero field with the model of point-like 
impurity scattering with a single phase shift. It is possible that the deficiencies
of the present theory may be due to this over-simplified model. Similar
difficulties in explaining microwave conductivity measurements in very
clean YBCO have led to consideration of impurity potentials with finite range.
This gives rise to considerable changes in the density of states at low 
frequencies due to renormalization of the d-wave gap. \cite{Priv} It seems
possible that such effects could lead to a better description of the $T$ 
dependence of $\kappa/T$ in zero field in the Born approximation. On the 
other hand we believe that the theory of Andreev scattering contained in the
theory of $\kappa$ in Ref.~\onlinecite{Vek} need not be altered because it 
agrees with our theory where the Andreev scattering is evident from the 
form of the self energy in the Gorkov equations. 

      Our results for the unitary limit and scattering rates $\delta \sim 0.01$ 
and 0.1 also apply to spin triplet states with vertical or horizontal line nodes
in Sr$_2$RuO$_4\,$. Indeed, the zero field measurments at very low $T$ 
strongly suggest a phase shift close to $\pi/2\,$. \cite{Suzuki}

\bigskip
We thank K. Scharnberg for helpful discussions.
\newpage
\newpage
\begin{figure}
\centerline{\psfig{file=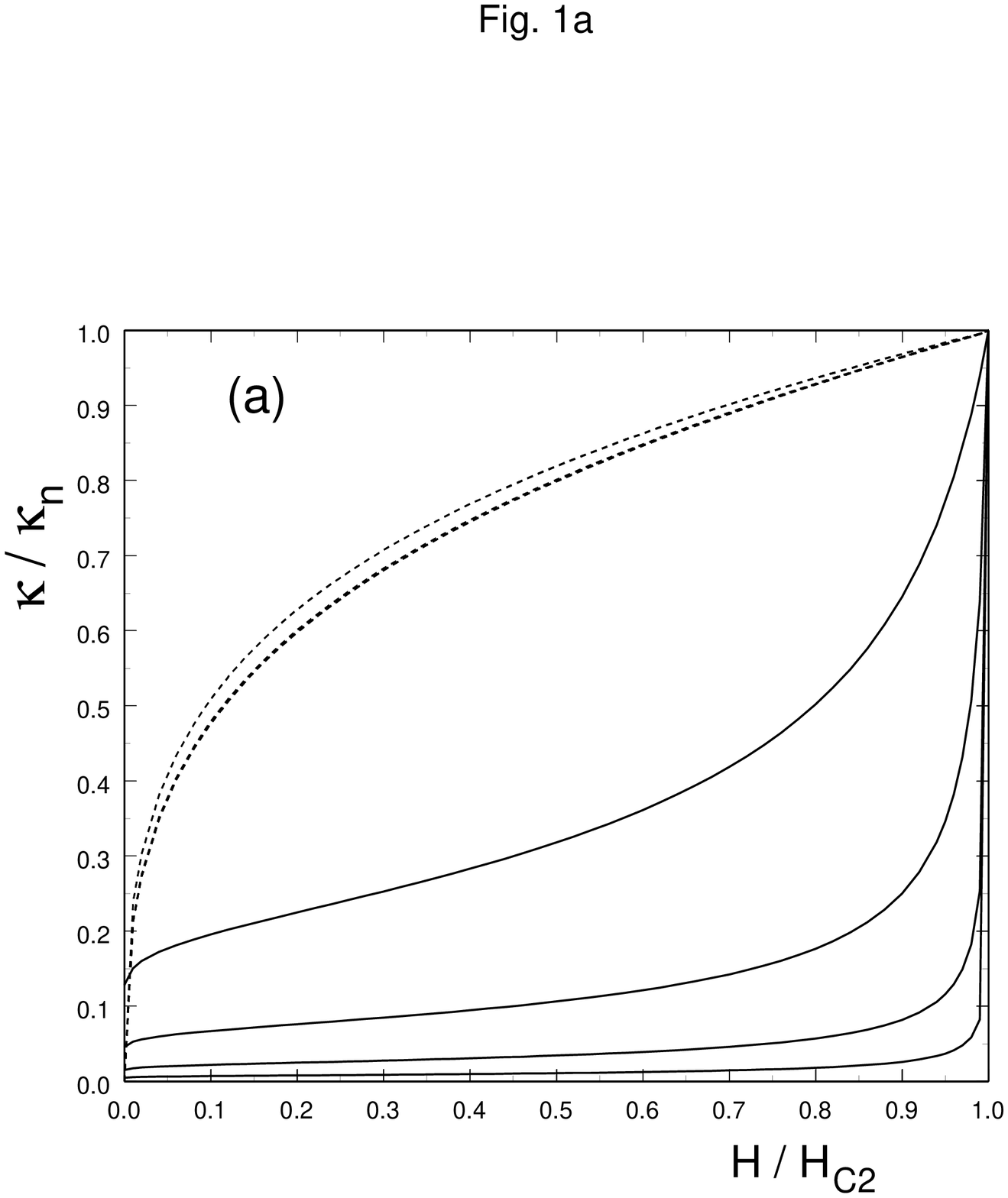,width=18cm,angle=0}}
\vskip -1cm
\caption{1a) Thermal conductivity ratio, $\kappa/\kappa_n\,$, vs $\,h=H/H_{c2}\,$ at 
$T=0$ in the Born approximation for the reduced scattering rates 
$\delta=\Gamma/\Delta_0=\,$ 0.1, 0.01, 0.001, and 0.0001 (solid curves,
from top to bottom) and the density of states at the Fermi energy, $N/N_0$ vs 
$h$ (dashed curves from top to bottom).}
\label{fig1a}
\end{figure}
\begin{figure}
\centerline{\psfig{file=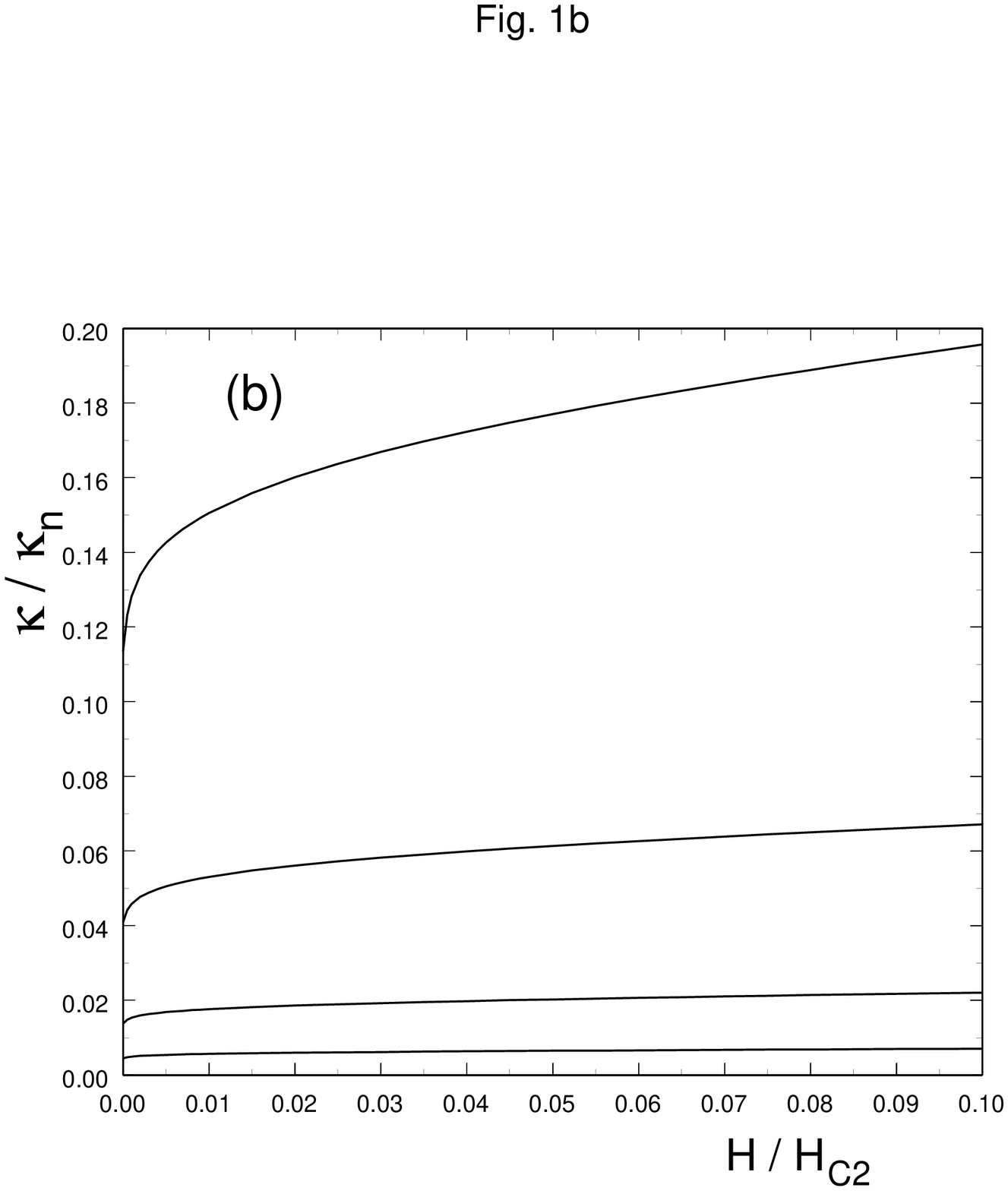,width=18cm,angle=0}}
\vskip -1cm
\caption{1b) The same as 1(a) but on a reduced scale.}
\label{fig1b}
\end{figure}
\begin{figure}
\centerline{\psfig{file=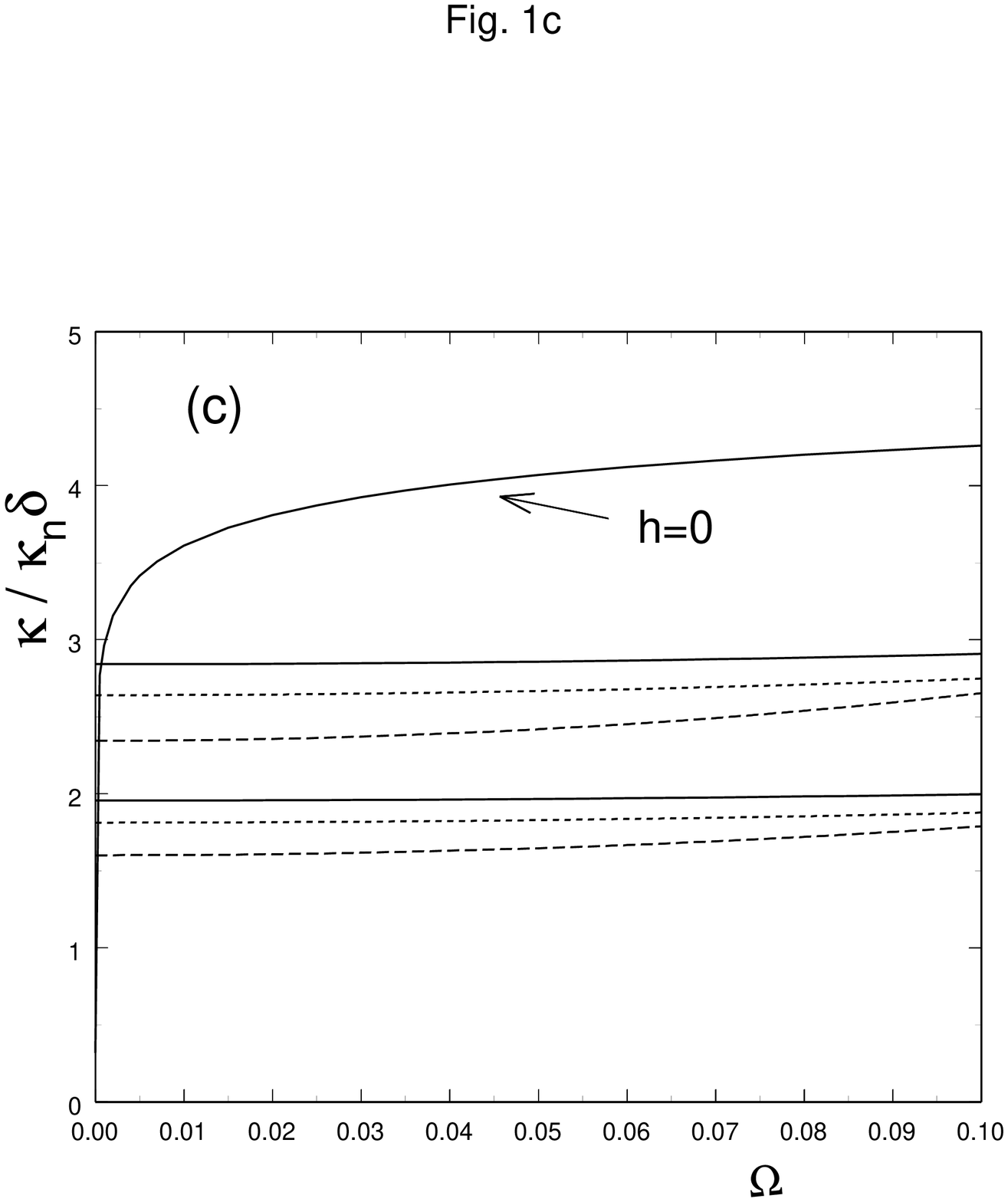,width=18cm,angle=0}}
\vskip -1cm
\caption{1c) The factor in the integrand of the $\omega$-integral for $\kappa\,$,
$\kappa(\omega)/\kappa_n\delta\,$, vs $\Omega=\omega/\Delta_0\,$. Upper
curves for $\delta=0.05$ in the Born approximation for $h=\,$ 0.1, 0.06, and 
0.02 and lower curves for $\delta=0.1$ and the same h (from top to bottom). 
Solid curve for $h=0$ and $\delta=0.1\,$.}
\label{fig1c}
\end{figure}
\begin{figure}
\centerline{\psfig{file=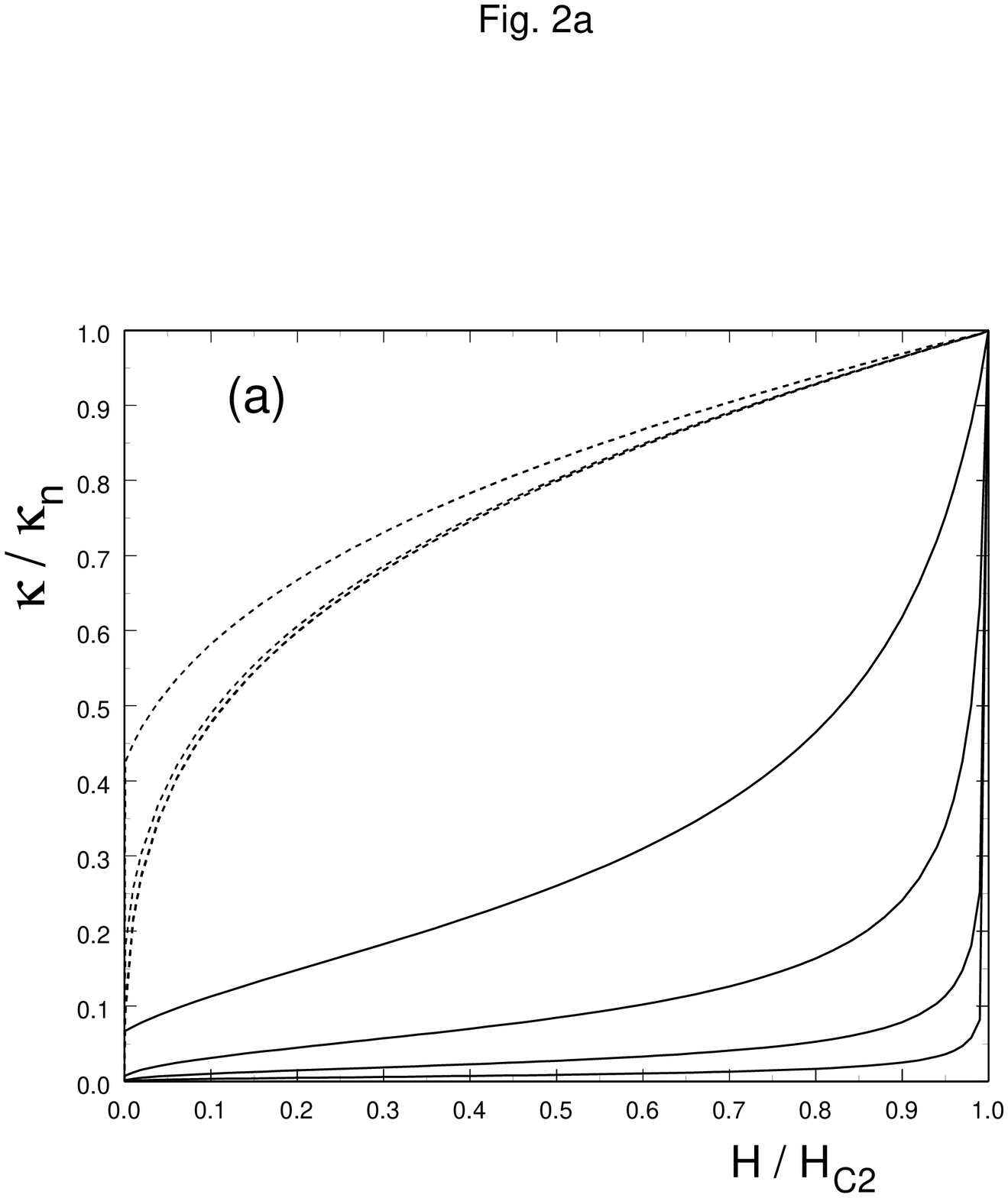,width=18cm,angle=0}}
\vskip -1cm
\caption{2a) $\kappa/\kappa_n$ vs $h$ at $T=0$ for impurity scattering phase shift
$\rho=0.495\pi$ and $\delta=\,$ 0.1, 0.01, 0.001, and 0.0001 (solid curves,
from top to bottom), and $N/N_0$ vs $h$ (dashed curves from top to bottom).}
\label{fig2a}
\end{figure}
\begin{figure}
\centerline{\psfig{file=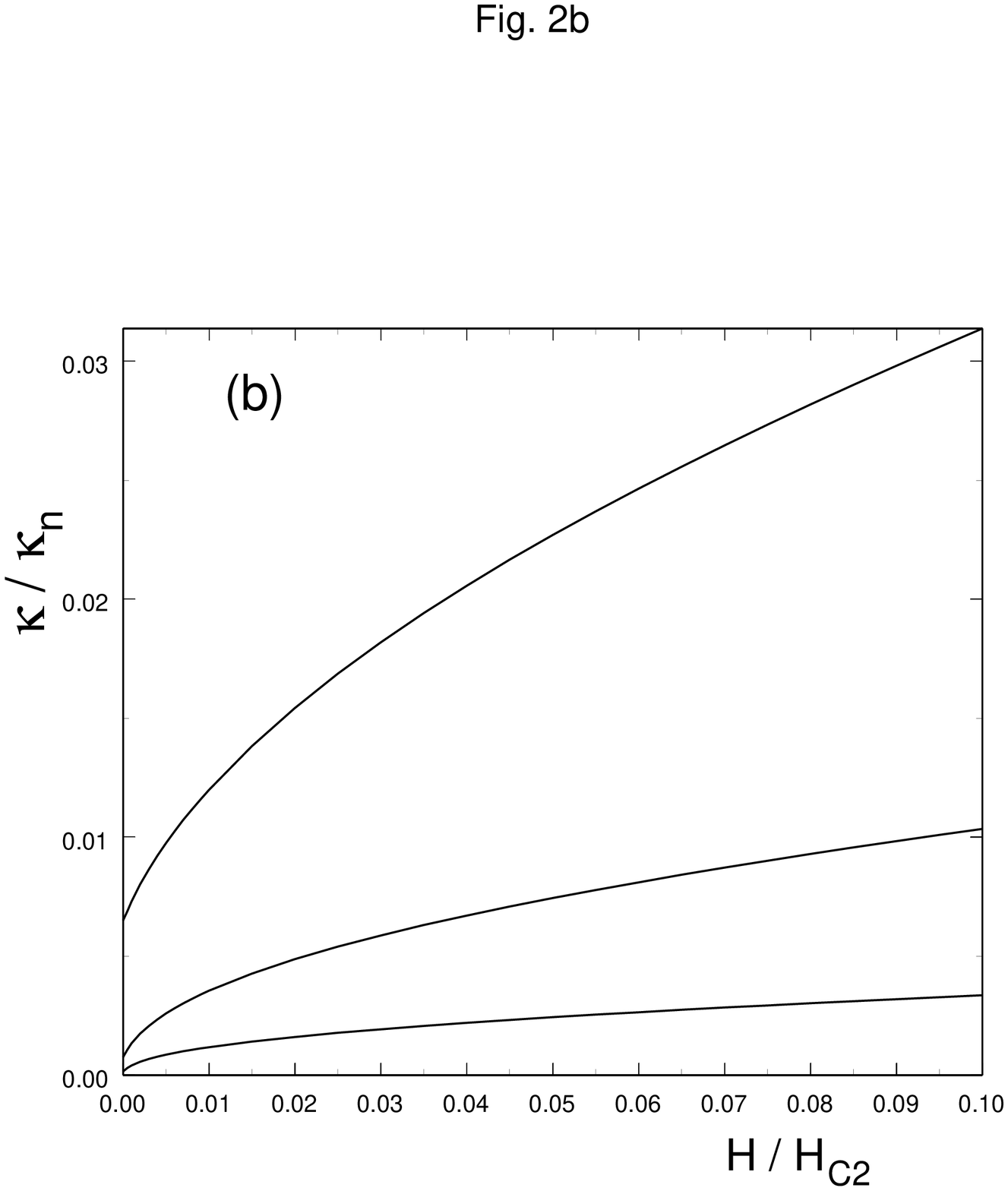,width=18cm,angle=0}}
\vskip -1cm
\caption{2b) The same as 2(a) but on a reduced scale for $\delta=0.01\,$,
0.001, and 0.0001.}
\label{fig2b}
\end{figure}
\begin{figure}
\centerline{\psfig{file=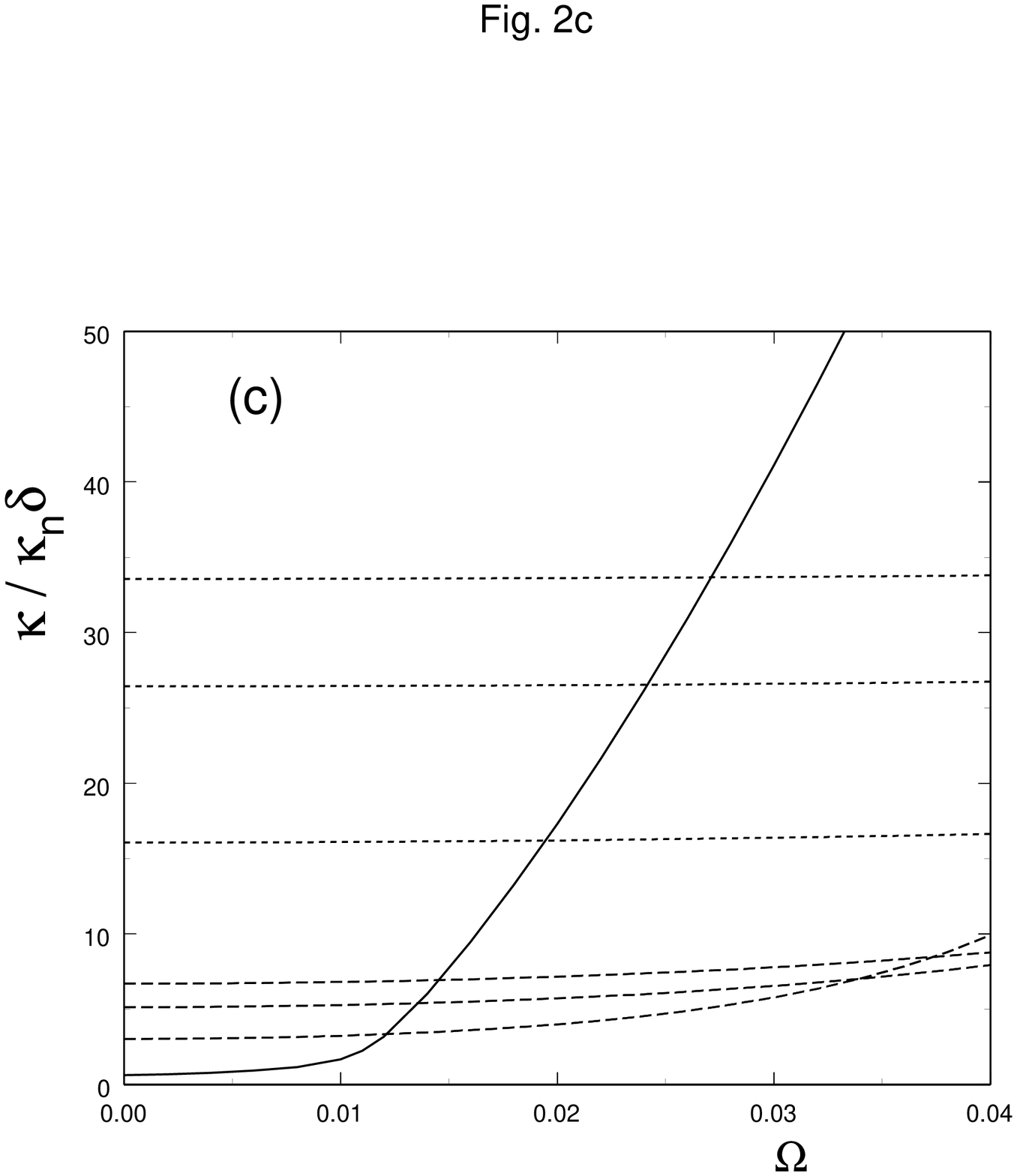,width=18cm,angle=0}}
\vskip -1cm
\caption{2c) $\kappa(\omega)/\kappa_n\delta\,$, vs $\Omega$ [see notation of Fig.1(c)]
for phase shift $\rho=0.495\pi$ and $\delta=0.0001\,$. Dashed curves for
$h=\,$ 0.1, 0.06, and 0.02 (from top to bottom). Upper curves for $\alpha=5$ 
and lower curves for $\alpha=31\,$. Solid curve for $h=0\,$.}
\label{fig2c}
\end{figure}
\begin{figure}
\centerline{\psfig{file=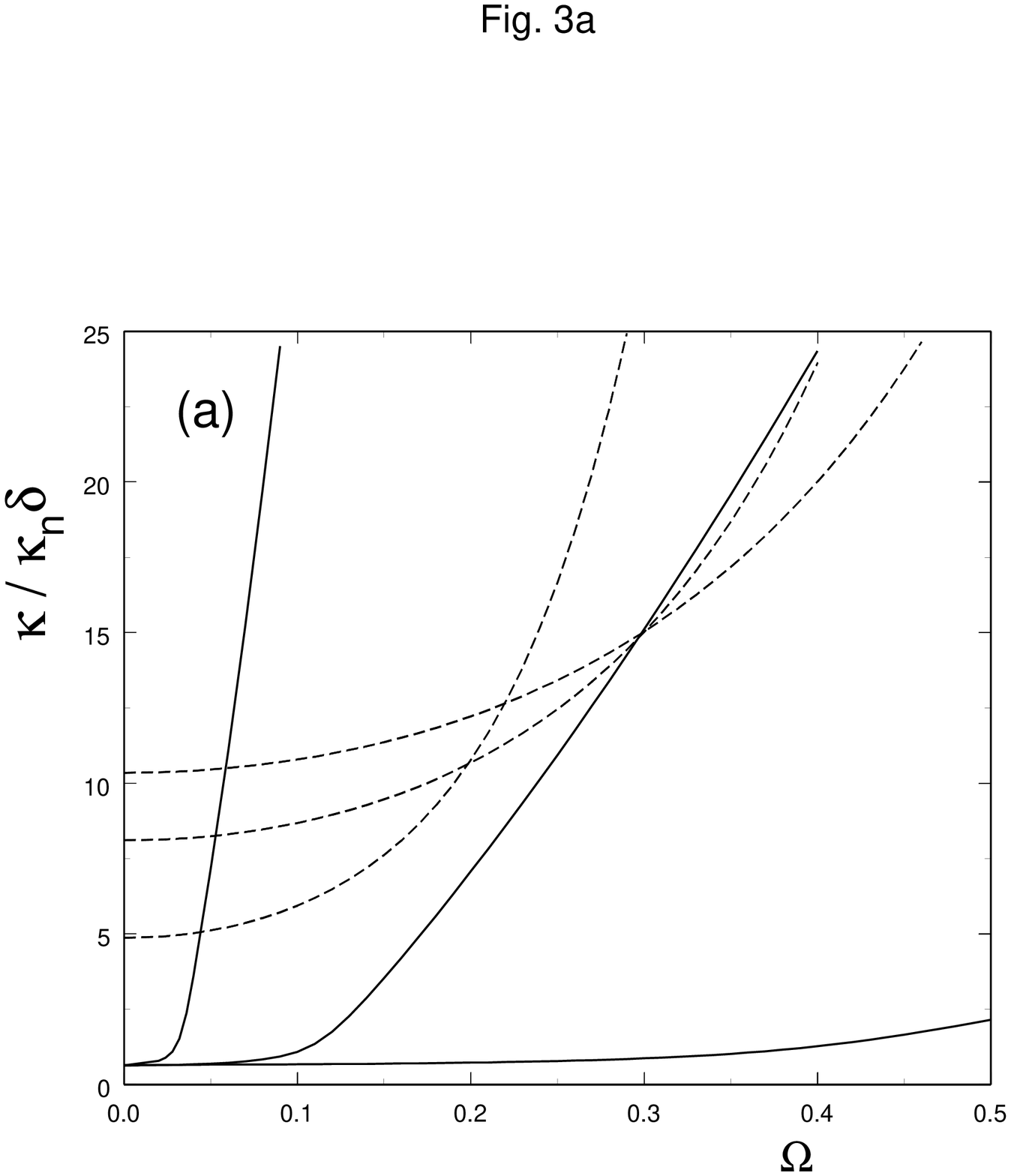,width=18cm,angle=0}}
\vskip -1cm
\caption{3a) Factor in the $\omega$-integral for $\kappa\,$, 
$\kappa(\omega)/\kappa_n\delta\,$, vs $\Omega=\omega/\Delta_0\,$, for 
phase shift $\rho=0.495\pi\,$. Solid curves for $h=0$ and $\delta=\,$ 0.001, 
0.01, and 0.1 (from left to right). Dashed curves for $\delta=0.001\,$,
$\alpha=5\,$, and $h=\,$ 0.1, 0.06, and 0.02 (from top to bottom)}
\label{fig3a}
\end{figure}
\begin{figure}
\centerline{\psfig{file=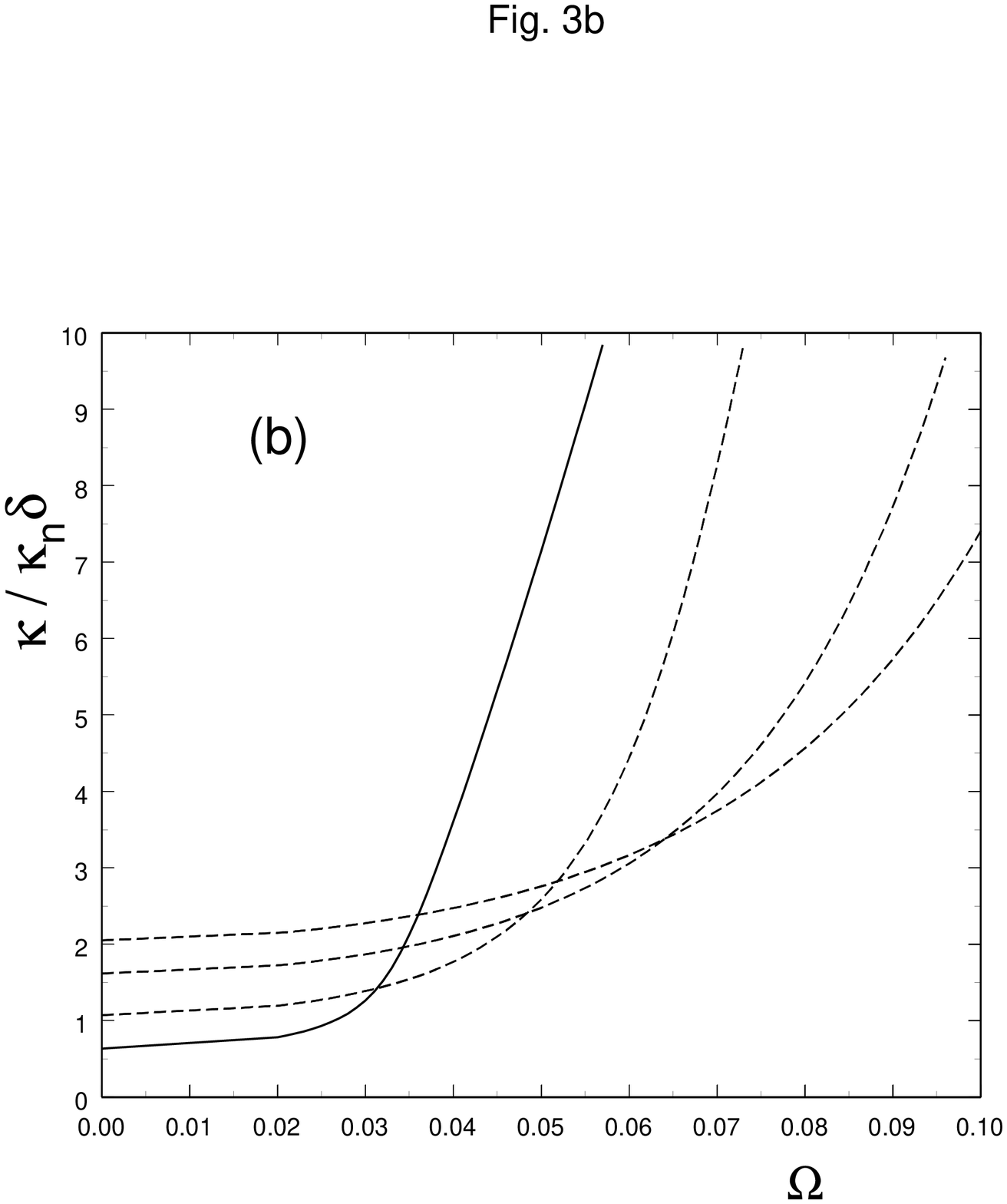,width=18cm,angle=0}}
\vskip -1cm
\caption{3b) Dashed curves for $\delta=0.001\,$, $\alpha=31\,$, and 
$h=\,$ 0.1, 0.06, and 0.02 (from top to bottom). Solid curve for $h=0\,$.}
\label{fig3b}
\end{figure}
\end{document}